\documentclass[sn-vancouver,Numbered]{sn-jnl}

\usepackage{graphicx}%
\usepackage{multirow}%
\usepackage{amsmath,amssymb,amsfonts}%
\usepackage{amsthm}%
\usepackage{mathrsfs}%
\usepackage[title]{appendix}%
\usepackage{xcolor}%
\usepackage{textcomp}%
\usepackage{manyfoot}%
\usepackage{booktabs}%
\usepackage{algorithm}%
\usepackage{algorithmicx}%
\usepackage{algpseudocode}%
\usepackage{listings}%

\begin{document}

\title[Article Title]{10 GHz Robust polarization modulation towards high-speed satellite-based quantum communication}


\author[1]{\fnm{Zexu} \sur{Wang}}\equalcont{These authors contributed equally to this work.}

\author[2,3]{\fnm{Huaxing} \sur{Xu}}\equalcont{These authors contributed equally to this work.}

\author[1]{\fnm{Ju} \sur{Li}}

\author[4,5]{\fnm{Jinquan} \sur{Huang}}

\author[6]{\fnm{Hui} \sur{Han}}

\author[2,3]{\fnm{Changlei} \sur{Wang}}

\author[2,3]{\fnm{Ping} \sur{Zhang}}

\author[1]{\fnm{Feifei} \sur{Yin}}

\author[1]{\fnm{Kun} \sur{Xu}}

\author*[4]{\fnm{Bo} \sur{Liu}}\email{liubo08@nudt.edu.cn}

\author*[1]{\fnm{Yitang} \sur{Dai}}\email{ytdai@bupt.edu.cn}

\affil[1]{\orgdiv{State Key Laboratory of Information Photonics and Optical Communications}, \orgname{Beijing University of Posts and Telecommunications}, \orgaddress{\city{Beijing}, \postcode{100876}, \country{China}}}

\affil[2]{\orgdiv{National Engineering Research Center for Public Safety Risk Perception and Control by Big Data}, \orgname{China Academy of Electronics and Information Technology}, \orgaddress{\city{Beijing}, \postcode{100041}, \country{China}}}

\affil[3]{\orgdiv{CETC Academy of Electronics and Information Technology Group Co., Ltd.},  \orgaddress{\city{Beijing}, \postcode{100041}, \country{China}}}

\affil[4]{\orgdiv{College of Advanced Interdisciplinary Studies}, \orgname{National University of Defense Technology}, \orgaddress{\city{Changsha}, \postcode{410073}, \country{China}}}

\affil[5]{\orgdiv{School of Electronics and Communication Engineering}, \orgname{Sun Yat-sen University, Shenzhen}, \orgaddress{\city{Guangdong}, \postcode{518107}, \country{China}}}

\affil[6]{\orgdiv{College of Computer}, \orgname{National University of Defense Technology}, \orgaddress{\city{Changsha}, \postcode{4100730}, \country{China}}}


\abstract{In practical satellite-based quantum key distribution (QKD) systems, the preparation and transmission of polarization-encoding photons suffer from complex environmental effects and high channel-loss. Consequently, the hinge to enhancing the secure key rate (SKR) lies in achieving robust, low-error and high-speed polarization modulation. Although the schemes that realize self-compensation exhibit remarkable robustness. Their modulation speed is constrained to approximately 2 GHz to avoid the interaction between the electrical signal and the reverse optical pulses. Here we utilize the non-reciprocity of the lithium niobate modulators and eliminate the modulation on the reverse optical pulses. As this characteristic is widely available in the radio-frequency band, the modulation speed is no longer limited by the self-compensating optics and can be further increased. The measured average intrinsic QBER of the different polarization states at 10 GHz system repetition frequency is as low as 0.53$\%$ over 10 min without any compensation. And the experiment simulation shows that the proposed scheme extends the transmission distance to more than 350 km. Our work can be be efficient performed to  the high-speed and high-loss satellite-based quantum communication scenario.}




\maketitle

\section{Introduction}\label{sec1}

Quantum key distribution (QKD) can provide information-theoretical-secure keys between distant communication parties, which are guaranteed by the quantum mechanics rule \cite{xu2020secure}. In the implementation of a global quantum network, satellite-based polarization-encoding QKD is preferred over optical fiber solution because of the negligible photon loss and decoherence experienced in empty space \cite{2017Satellite}. The enhancement of its secure key rate (SKR) faces three challenges in the photonic layer: the generation, preparation and detection of photons. The generation of photons with high repetition-rate becomes easy with the development of light sources. During satellite-to-ground communication, the number of photons arriving the receiver is significantly reduced due to channel loss. Therefore, although the efficiency of the
detector still has an impact, it is no longer the main 
obstruct limiting the key generation rate. Meanwhile, due to the complex environmental effects and limited loads in the satellite, the speed and robustness of the photons preparation become the bottleneck to the improvement of the SKR.

The schemes currently available are unable to guarantee robustness and low quantum bit error (QBER) with high-speed polarization modulation. The previous researchers adopt the setups that use an inline phase modulator \cite{grunenfelder2018simple} or Mach-Zehnder interferometer \cite{li2023high}. An experiment with a 5 GHz repetition frequency was reported in 2020 \cite{grunenfelder2020performance}. However, it has more drawbacks when obtaining a higher modulation speed, such as severe polarization mode dispersion (PMD) and a half-wave voltage of up to 7 V. Moreover, their optical path can easily be affected by the environment. To solve this problem, schemes based on the Sagnac \cite{xu2020intrinsic} or Faraday-Michelson interferometer \cite{stein2023robust} are carried out. Their self-compensation optics can achieve excellent system robustness and low QBER modulation \cite{agnesi2019all}. However, there is interaction between electrical signals and optical pulses that propagate in the opposite direction in the phase modulator (PM). To avoid the impact of this reverse modulation, it is necessary to ensure that only forward pulses are modulated. The threshold of the modulation speed is 2.27 GHz when using commercial components \cite{luo2022intrinsically}. It can be increased by shortening the waveguide length. However, this approach will cause significant difficulties for electric driving, and the custom phase modulator also raises the overall cost.

\begin{figure}[ht]
\centering
\includegraphics[width=10cm]{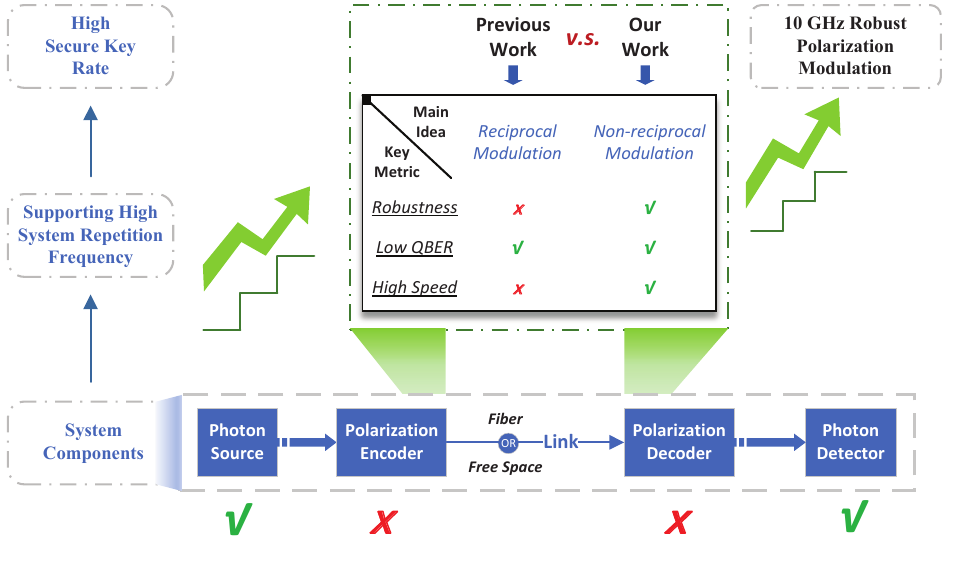}
\caption{The key metric of QKD is the secure key rate (SKR). And the improvement of SKR can be started from the system repetition frequency, which needs us to achieve robust, low-error and high-speed modulation.}
\label{fig1}
\end{figure}

In this paper, we removes the severe interaction between electrical signals and reverse optical pulses in the modulator by using the non-reciprocal characteristic of lithium niobate (LiNbO$_3$) modulators. The modulation speed therefore is no longer limited by the optical path, while the system retains the high robustness of the self-compensating optics. This allows us to break the 2.27 GHz threshold using only commercial component parts. More importantly, since the non-reciprocity is widespread in the radio-frequency (RF) band, the enhancement of system repetition frequency only needs an ultrashort laser with high repetition-rate and appropriate circuit to perform the modulation. This increase in system repetition frequency directly corresponds to a high SKR. The measured average intrinsic quantum bit error rate of different polarization states is as low as 0.53$\%$ over 10 min at the system repetition frequency of 10 GHz without any compensation. The simulation results prove that the proposed scheme achieves a higher SKR under the same distance condition and extends the transmission distance to more than 350 km. Our work that provides robust, low-error and high-speed polarization modulation for satellite-based quantum communications can maintain a high level of communication security over long distances.

\section{Methods}\label{sec2}

\begin{figure}[ht]
\centering
\includegraphics[width=10cm]{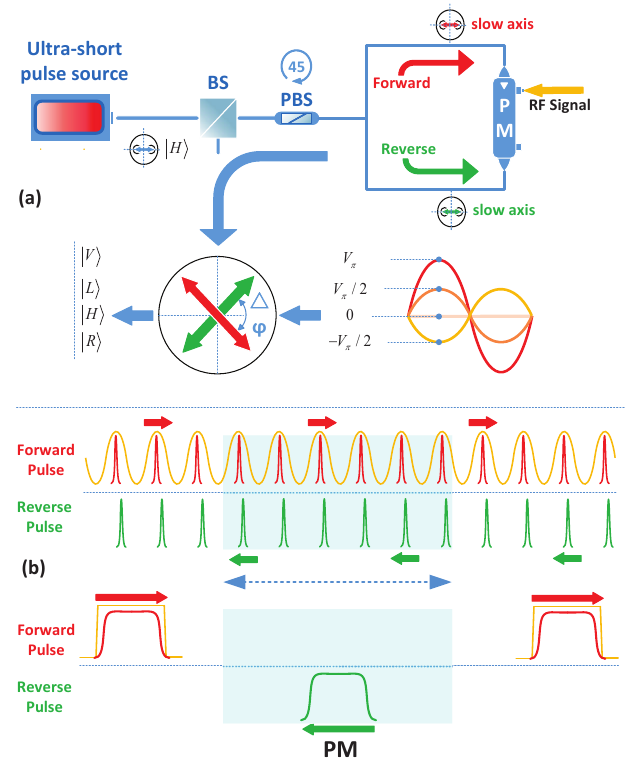}
\caption{(a) The schematic of polarization modulation. Ultrashort optical pulses are transmitted into the polarization encoder which is consist of a PM and a 45-degree PBS. An RF sine wave is loaded onto the PM as the electrical signal to realize polarization-encoding. Four different modulation voltages correspond to four required polarization states. (b) The distribution of the optical pulse. The forward optical pulses are plotted in red, and the reverse are plotted in green. The electrical signals loaded on the PM are plotted in yellow and propagate forward. PBS, polarization beam splitter; BS, beam splitter; PM, phase modulator.}
\label{fig2}
\end{figure}

The principle of robust, low-error, and high-speed polarization modulation in the proposed scheme is illustrated in Fig.~\ref{fig2}. As shown in Fig.~\ref{fig2}(a), high repetition-rate ultrashort pulses are transmitted into the beam splitter (BS) and are divided into two parts. The reflected part can be used for monitoring, and the transmitted part enters the polarization encoder for modulation. The polarization encoder is a Sagnac interferometer consisting of a LiNbO$_3$-PM and a 45-degree polarization beam splitter (PBS). The optical pulses entering the PM in both the forward and reverse directions are transmitted in the slow axis of the polarization-maintaining fiber (PMF) and undergo the same optical path. Therefore, they achieve self-compensation from the impact of the phase noise introduced by the transmission medium and avoid PMD. This guarantees the robustness of the system. The polarization state of the modulated optical pulses can be expressed as
\begin{equation}
|\psi_{out}\rangle =[|D\rangle+\text{exp}(i\pi V(t)/V_{\pi F})|A\rangle]/\sqrt{2}
\label{eq1}
\end{equation}
where $V(t)$ is the electrical signal loaded on the PM and $V_{\pi F}$ is the half-wave voltage for forward modulation. Because of the nonreciprocal characteristic of LiNbO$_3$ modulators, only the forward process of modulation is considered in Eq.~\ref{eq1}. We select the electrical signal as $V(t)=V_0 \sin(2\pi f_dt)$. In addition, $f_d$ should be an integer multiple of the repetition-rate of optical pulses. To achieve optimal modulation, we need to adjust the delay such that the peak of the optical pulses and the electrical signal coincide in the time domain. Synchronization should also be ensured. The output polarization state is $|H\rangle$ when $V_0$ is not applied. If $V_0$ is set to $V_{\pi F}/2$, $-V_{\pi F}/2$, they correspond to the polarization states of $|L\rangle$, $|R\rangle$. Alternatively, if $V_0$ is set to $V_{\pi F}$, the output is $|V\rangle$. These two sets of orthogonal polarization states can serve as unbiased bases for the BB84 protocol.

Figure~\ref{fig2}(b) focuses on the optical pulse distribution within the Sagnac interferometer, which clearly shows an improvement in the modulation speed compared with previous schemes. In our scheme, more optical pulses are simultaneously modulated by an RF sine wave at the same time. Both forward and reverse optical pulses exist in the PM, and there is no need to compensate for reverse modulation. In contrast, the previous scheme shows that only one optical pulse is modulated by a square wave simultaneously which represents a lower modulation speed. To avoid interaction between the electrical signal and the reverse optical pulses, the previous scheme must make them enter the PM at different times. Because of the impracticable compensation of reverse modulation, it is necessary to further shorten the length of the modulator waveguide to separate the optical pulses in the time domain when the modulation speed increases. The chain between the length of the waveguide and the modulation speed limits the system repetition frequency to: $f_{max}=c/[c(T_0+T_e )+2n_0 L]$, where $T_0$ is the width of the optical pulses; $T_e$ is the duration of the square wave; $L$ is the length of the PM waveguide; $n_0$ is the refractive index of the LiNbO$_3$ crystal.

The implementation of our robust, low-error, and high-speed polarization modulation is based on the non-reciprocal characteristic of LiNbO$_3$ modulators; only the forward light pulses are modulated, whereas the phase of the reverse pulses can be considered to be unaffected by the electrical signal. By utilizing this characteristic, our scheme breaks through the chain between the modulation speed and the waveguide length. It can serve as the encoder or the decoder, when the satellite-based QKD systems with high repetition frequency is needed to meet the demand for improved SKR. This improvement can be easily applied in all schemes with bidirectional optical path setup, including phase \cite{muller1997plug} and time-bin \cite{boaron2018simple} encoding.

The LiNbO$_3$-PM is a traveling-wave modulation device. It is available for both the forward and the reverse modulation. Specifically, in the Sagnac loop, the pulses entering the PM in the forward direction are transmitted in the same direction as the electrical signal, while the pulses entering in the reverse direction are transmitted in the opposite direction. However, the mechanisms by which electrical signals modulate forward and reverse pulses are different. Here, we use the half-wave voltage to measure the modulation performance of the optical pulses. In LiNbO$_3$-PM, the forward half-wave voltage $V_{\pi F}$ can be considered constant when the velocities of the optical and modulation fields are well matched \cite{li2006novel}. The expression for the reverse half-wave voltage $V_{\pi R}$ is
\begin{equation}
V_{\pi R}(f)=V_{\pi F}|\frac{2\pi f\tau_d}{\sin(2\pi f\tau_d)}|
\label{eq2}
\end{equation}
where $f$ is the frequency of the electrical signal and $\tau_d$ is the average propagation time of the optical field and the modulation fields. 

\begin{figure}[ht]
\centering
\includegraphics[width=11cm]{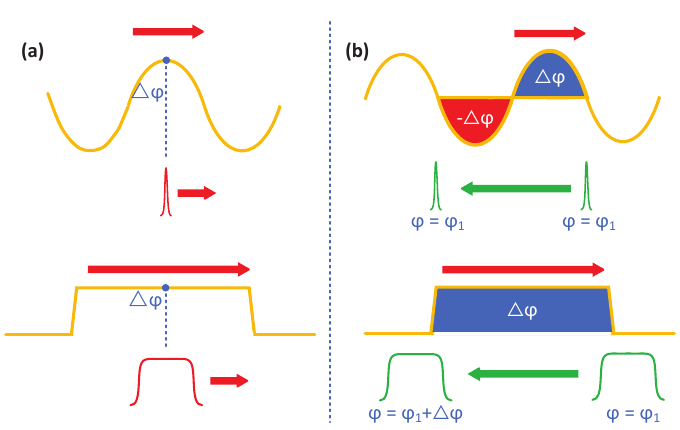}
\caption{The schematic diagram of the modulation process. (a) The result of forward modulation is depended on the corresponding modulation voltage. (b) The result of reverse modulation is depended on the integration of modulation voltage. The previous scheme applies the square wave which causes obviously phase shift on the reverse pulses. However, our solution uses RF sin wave instead so the overall modulation effect can be ignored. The forward pulses are plotted in red. The reverse pulses are plotted in green. And the electrical signals loaded on the PM are plotted in yellow.}
\label{fig3}
\end{figure}

A visual explanation is shown. Figure~\ref{fig3}(a) shows that when the optical pulse and the electrical signal propagate in the same direction, the interaction between them is a continuous process. The relative positions of the electrical signal and the optical pulse remain constant when their velocities are matched. The result is only dependent on the corresponding voltage for modulation. In contrast, the interaction between the optical pulse and the electrical signal is very short due to the opposite propagation direction. The entire phase modulation can be considered an integration process. Thus, the performance of reverse modulation is related to the result of integration. Therefore, we can explain the different performances of the RF sine wave (our work) and square wave (previous work) during the reverse modulation process in Fig.~\ref{fig3}(b). The modulation voltage integration is zero within a period for the sine wave, so that the effects of the positive and negative voltages nearly cancel out. The overall modulation effect can be ignored. However, the modulation voltage integration is never zero for the square wave. A duration of a few hundred picoseconds can cause a significant reverse modulation.

\begin{figure}[ht]
\centering
\includegraphics[width=10cm]{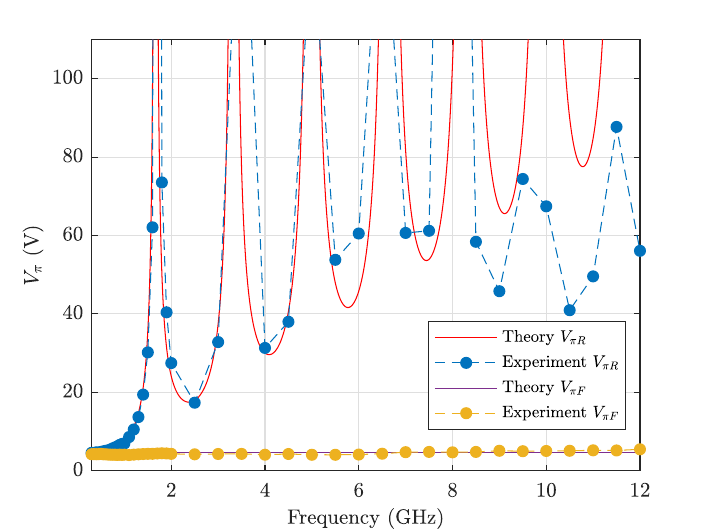}
\caption{The experimental and theoretical data for forward and reverse half-wave voltages. The forward half-wave voltages can be regarded as a constant in the measurement range. Meanwhile, a strong frequency correlation exists for the reverse half-wave voltage. The non-reciprocal characteristic of LiNbO$_3$-PM is widely present in the RF band and becomes more pronounced with increasing frequency.}
\label{fig4}
\end{figure}

As Eq.~\ref{eq2}, a strong frequency correlation exists for $V_{\pi R}$. We measured the forward and reverse half-wave voltages to confirm it. The Frequency$-V_\pi$ curves in Fig.~\ref{fig4} are plotted. It shows that $V_{\pi R} \approx V_{\pi F}$ when the modulation frequency is below 1 GHz; $V_{\pi R}$=8.6 V $\approx 2V_{\pi F}$ when the frequency increases to 1.1 GHz. Afterwards, the frequency continues to increase to 10 GHz, with $V_{\pi R}$=67.4 V $\approx 13.5V_{\pi F}$. There is a slight increase in $V_{\pi F}$ during the measurement with the frequency variation. This is because the bandwidth of the modulator is limited, as opposed to being ideal. However, $V_{\pi R}$ varies significantly with frequency, particularly at $f=N/2\tau_d$, where it has the local maximum. We get $V_{\pi R}$=781.4 V which is the highest within the measurement range. At this point, if the same voltage is used for phase modulation, the forward pulses have a phase shift of 1 rad, and the reverse pulses have a phase shift of only 0.006 rad. 

Moreover, we did not cause a rapid increase in the equivalent half-wave voltage compared to the previous approach. The equivalent half-wave voltage of our encoder is generally consistent with the half-wave voltage of the PM in the RF band. The equivalent half-wave voltage is 5.47 V at 10 GHz, and the half-wave voltage of the PM at the same point is 5.06 V. This is an increase of only 8$\%$.

Figure~\ref{fig4} shows that the non-reciprocity is widespread in the RF band and becomes more pronounced with increasing frequency. Further improvement in the modulation speed can be easily achieved by using the laser with high repetition-rate and the matching modulation circuits. We just carried out an experiment under 10 GHz in this work, which is sufficient to demonstrate the ability of our robust encoder in supporting high repetition frequency. In addition, 10 Gbps is a commonly used bit rate in classical communication. The matching ultrashort pulse source and RF circuit are easy to implement. Our scheme represents an important step for satellite-based quantum communication in catching up with classical communication.

\section{Results}\label{sec3}

\begin{figure}[ht]
\centering
\includegraphics[width=12cm]{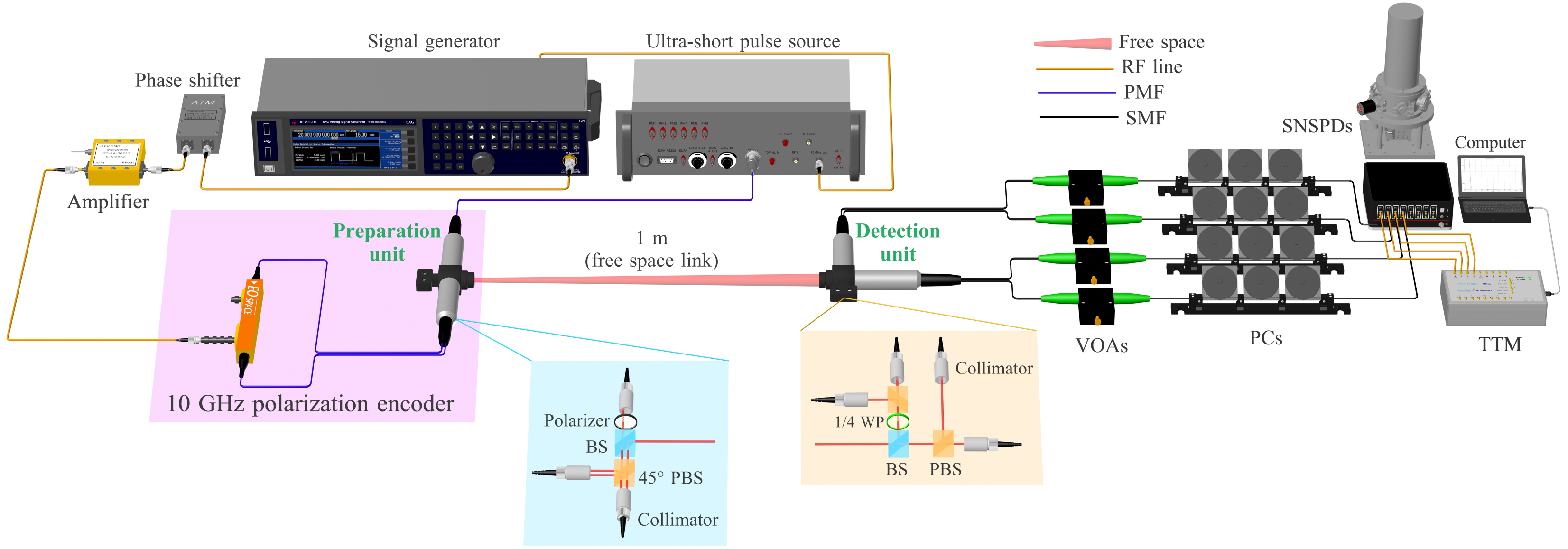}
\caption{The experimental setup. VOAs, variable optical attenuators; PCs, polarization controllers; SNSPDs, superconducting-nanowire single-photon detectors; TTM, time tagger module; PBS, polarization beam splitter; BS, beam splitter; PM, phase modulator; WP, wave plat; PMF, polarization-maintaining fiber; SMF, single mode fiber.}
\label{fig7}
\end{figure}

We conducted an experiment based on the setup shown in Fig.~\ref{fig7}, and evaluated the polarization extinction ratio (PER) and the QBER of four prepared polarization states. The light source of the system is an optical frequency-comb laser developed in our laboratory. The optical pulse output by the laser with a 10 GHz repetition-rate can achieve a full width at half maximum (FWHM) of approximately 2.16 ps after using a wave-shaper for spectral shaping and dispersion compensation. The application of an ultrashort pulse source is necessary for our proposed scheme. The impact of the pulse width on the performance of the system is discussed in the next section.

The components in the preparation unit, other than the PM, are integrated into a three-port device. The first port corresponds to the input of the collimator and receives pulses generated by the laser. The second port corresponds to two polarization-maintaining fiber tails of a 45-degree PBS, which are connected to a PM (Eospace, PM-5V4-40-PFU-PFU-UV) in forward and reverse directions. We use an analog signal generator (Keysight, EXG N5173B) to generate a 10 GHz sine wave. The signal generator is synchronized with the laser through a 10 MHz clock. To achieve optimal modulation, a phase shifter is used to change the phase of the electrical signal (equivalent to adjusting the delay) such that the peaks of the electrical signal and the optical pulses match in time. The electrical signal is loaded onto the PM after it is amplified by an amplifier to reach the required voltage for the four required polarization states. The third port outputs the prepared polarization states in free space.

\begin{figure}[ht]
\centering
\includegraphics[width=12cm]{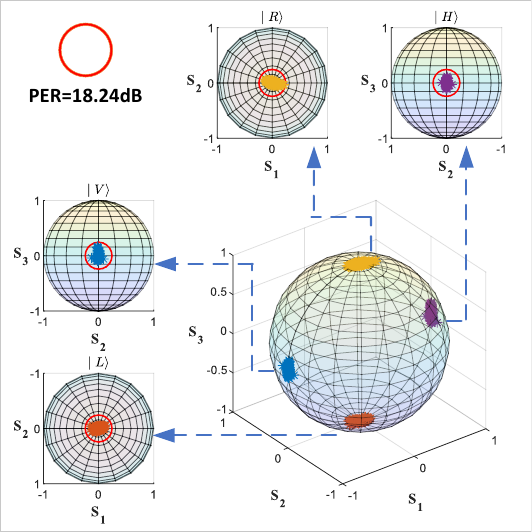}
\caption{Four prepared polarization states received by the polarization analyzer over 1 h 25 min without any compensation. The threshold of the PER is in the northwest corner. It corresponds to approximately 1.5$\%$ of QBER. We also mark it in Poincar{\'e} spheres.}
\label{fig8}
\end{figure}

In addition, we integrated the detection unit into a three-port device. The first port receives polarized light from the preparation unit, and the four outputs of the second and third ports decay to the single-photon level after equal loss by variable optical attenuators. We then adjusted the polarization controllers to effectively receive the prepared polarization states by superconducting nanowire single-photon detectors. The detection signals collected by the channels are fed back to a time-correlated single-photon-counting device (Swedish instruments, Time Tagger Ultra). Subsequently, the data are processed by a computer to obtain the QBER.

First, we evaluated the performance of the prepared polarization states using the classical optical method. A free-space polarization analyzer (Thorlabs, PAX 1000) is applied to analyze the output of the preparation unit. The polarization analyzer records the normalized Stokes vectors and plots them in Poincar{\'e} spheres. Polarization states with a polarization extinction ratio of 18.24 dB are also marked. This corresponds to approximately 1.5$\%$ of the QBER. Figure 5 shows that the PER of the four polarization states exceed 18.24 dB (lower than 1.5$\%$ QBER) over 1 h 25 min without any compensation.

\begin{figure}[ht]
\centering
\includegraphics[width=10cm]{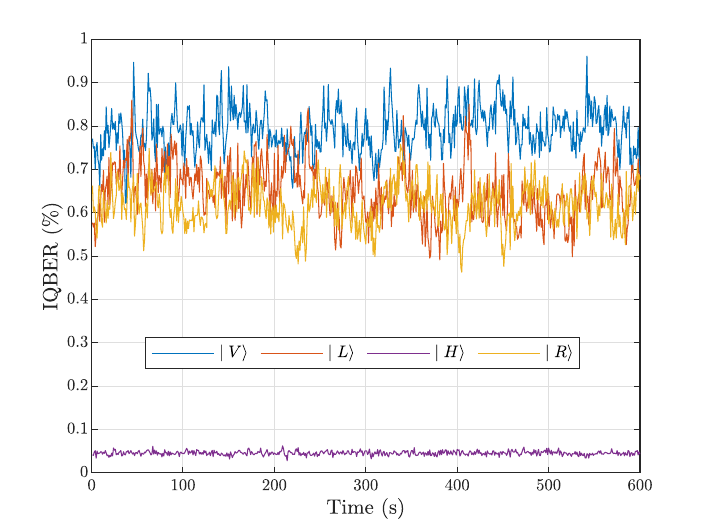}
\caption{The measured IQBERs of $|H\rangle$ (purple), $|L\rangle$ (red), $|V\rangle$ (blue) and $|R\rangle$ (yellow). The average IQBER of different polarization states is as low as 0.53$\%$ over 10 min without any compensation.}
\label{fig9}
\end{figure}

Second, we followed the setup shown in Fig.~\ref{fig7} and conducted a single-photon level test. The preparation unit generates optical pulses with a fixed polarization state during each measurement cycle and continuously transmits them to the detection unit. To characterize the purity and robustness of each polarization state prepared by our polarization modulation scheme, the intrinsic quantum bit error rate (IQBER) is used as the core indicator to measure the effectiveness and is defined as: IQBER$=(C_B^\bot-D_B^\bot )/[C_B^\bot-D_B^\bot+C_A-D_A ]$, where $C_A$ is the photon count of the polarization state as the preparation target; $D_A$ is the corresponding dark count; $C_B^\bot$ is the photon count of the polarization state orthogonal to the prepared polarization state; $D_B^\bot$ is the corresponding dark count \cite{stein2023robust}. Characterization of the polarization state modulated by the scheme is achieved by measuring IQBERs. This demonstrates the ability of the proposed scheme to generate the polarization states required for the BB84 protocol.

Figure~\ref{fig9} shows the result of each polarized state (repeated 1 s measurements over a 10 min time window). A positive correlation exists between the modulation voltage and average IQBER of the system. The generated polarization state $|H\rangle$ has the most stable and lowest average IQBER of 0.046$\%$ when no modulation voltage is applied. The average IQBER of the corresponding states $|L\rangle$ and $|R\rangle$ increases to 0.656$\%$ and 0.617$\%$, respectively, when the applied modulation voltages are $V_{\pi/2}$ and $-V_{\pi/2}$. The modulation voltage continues to increase to $V_\pi$. More phase noise is introduced. The average IQBER of state $|V\rangle$ is 0.798$\%$. In addition, we tested the polarization state at a modulation voltage of $3V_{\pi/2}$. The corresponding average IQBER is the highest at 0.959$\%$, when the modulation voltage is also the highest. There are two reasons for this deterioration. The fluctuation of the amplitude increases with a higher modulation voltage, so more phase noise is introduced. Furthermore, with the use of an RF sine wave, a higher modulation voltage causes the electrical signal to become more uneven. This also causes deterioration of the IQBER. Overall, the measured average intrinsic quantum bit error rate of the different polarization states is 0.53$\%$ over 10 min without any compensation.

\section{Discussion}\label{sec4}

We realized robust, low-error, and high-speed polarization modulation compared to the previous scheme. Two major changes are applied to support this improvement: optical pulses with a narrower pulse width and RF sine waves instead of square waves. Here, we continue to use the threshold (PER=18.24 dB, corresponding to QBER=1.5$\%$) to evaluate whether the system can achieve low-error polarization modulation. The impact of these changes on the system performance will be discussed. The proposed scheme is simulated to demonstrate its potential to significantly improve the performance of satellite-based quantum communications.

\begin{figure}[ht]
\centering
\includegraphics[width=10cm]{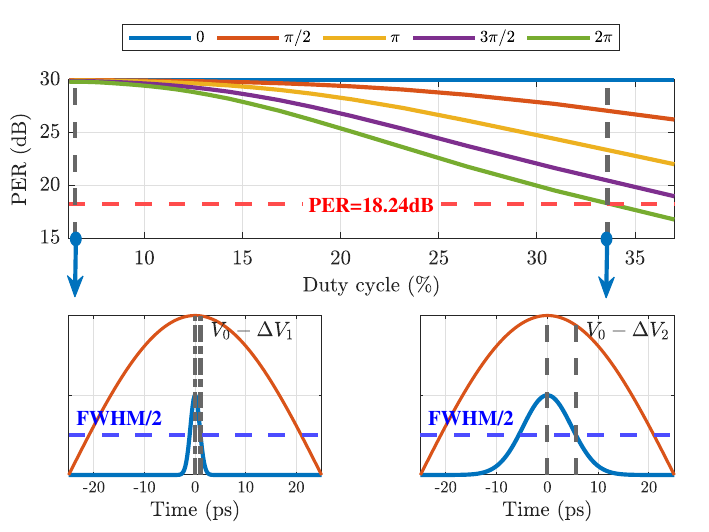}
\caption{The effect of the duty cycle on the polarization extinction ratio (PER). The photons of the optical pulse correspond to different modulation voltages because of the uneven envelope of the sine wave. Phase noise is introduced and makes the PER worse. The PER of the system is set to 30 dB in the simulation, when modulation voltage is not applied. The different colored curves represent different modulation phases applied in the above part. The relative position of optical pulse and electrical signal with the duty cycle of 6.48$\%$ and 33.8$\%$ is shown in the lower part. The RF sine wave is plotted in red while optical pulse is plotted in blue.}
\label{fig10}
\end{figure}

Our experimental setup applies an ultrashort pulse source, and the pulse width is reduced by $1 \thicksim 3$ orders of magnitude compared with the previous scheme. This change is necessary to achieve low-error and high-speed modulation. As shown in Fig.~\ref{fig10}, the peak of the optical pulse is also modulated by the peak voltage $V_0$ of the electrical signal. Because the envelope of the electrical signal is a sine wave, the peak is not as flat as that of the square wave. The photons of the optical pulse correspond to different modulation voltages. For example, photons located at half of the FWHM have a modulation voltage of $V_0-\Delta V$. The difference $\Delta V$ between the ideal and actual modulation voltages directly introduces phase noise, and thus worsens the PER of the system.

We use the duty cycle, which represents the ratio of the FWHM of the optical pulse and electrical signal, to reach a general conclusion and conduct a simulation. Figure~\ref{fig10} shows that, as the duty cycle increases, the PER of the system gradually deteriorates. At a duty cycle of 6.48$\%$ (which corresponds to the optical pulse with 2.16 ps pulse width and 10 GHz sine wave used in the experiment), $\Delta V_1=0.0023V_0$. When the duty cycle is 33.8$\%$, $\Delta V_2=0.062V_0$. The PER will be lower than the threshold if the duty cycle exceeds 33.8$\%$. This represents a poor performance in the modulation error. Therefore, a wider pulse width implies a higher duty cycle, which leads to a worse PER when the shape of the electrical signal is constant in practical QKD systems. 

\begin{figure}[ht]
\centering
\includegraphics[width=10cm]{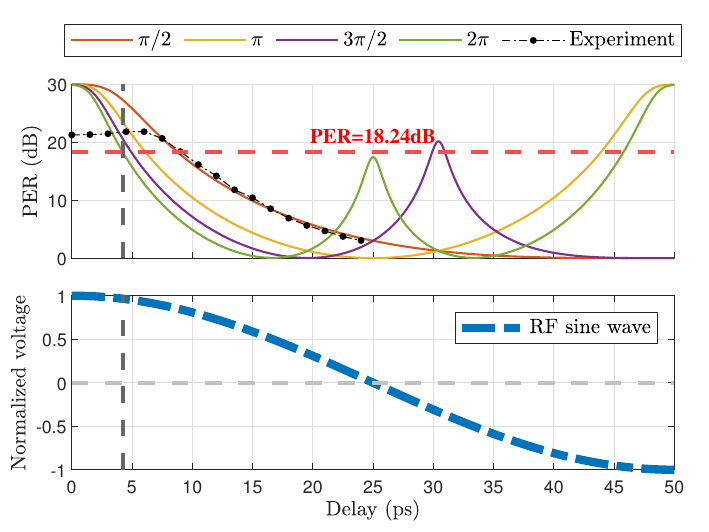}
\caption{The effect of the internal synchronization on the polarization extinction ratio (PER). The sine wave is not flat relative to the square wave. The modulation voltage, corresponding to the peak of the optical pulses, is no longer optimal when a delay occurs. This leads to a poor PER. The delay between the optical pulses and the electrical signal should be controlled within 4.25 ps. The different colored curves represent different modulation phases applied in the above part.}
\label{fig11}
\end{figure}

Synchronization is an important step in the development of practical QKD systems. This step consists of two parts: internal synchronization among the components of the system, and external synchronization between the transmitter (Alice) and the receiver (Bob). In this work, we focus on the impact of internal synchronization. A 10 MHz clock is applied to synchronize the devices in the system, including the signal generator and laser. The peaks of the optical pulses and the electrical signal match in time, resulting in the lowest modulation error. However, in practical systems, the phase drift caused by the thermal noise accumulated during the long-term operation of electrical devices, as well as the jitter of the optical pulse repetition-rate, causes the peak of the electrical signal and ultrashort optical pulses to separate. The previous scheme has a relatively high tolerance for jitter caused by poor synchronization. The flat part of the square wave guarantees that even if a delay occurs between the optical pulses and electrical signal, the modulation voltage will not change. However, our scheme uses an RF sine wave as the electrical signal, which is not flat compared to a square wave. The corresponding modulation voltage is no longer the ideal voltage when a delay occurs, which leads to poor PER. Figure~\ref{fig11} shows the results for this issue. The delay between the optical pulses and the electrical signal should be within 4.25 ps with a duty cycle sufficiently small such that the PER is higher than the threshold we set. The experimental results are also shown in Fig.~\ref{fig11} and are consistent with the simulation. 

\begin{figure}[ht]
\centering
\includegraphics[width=10cm]{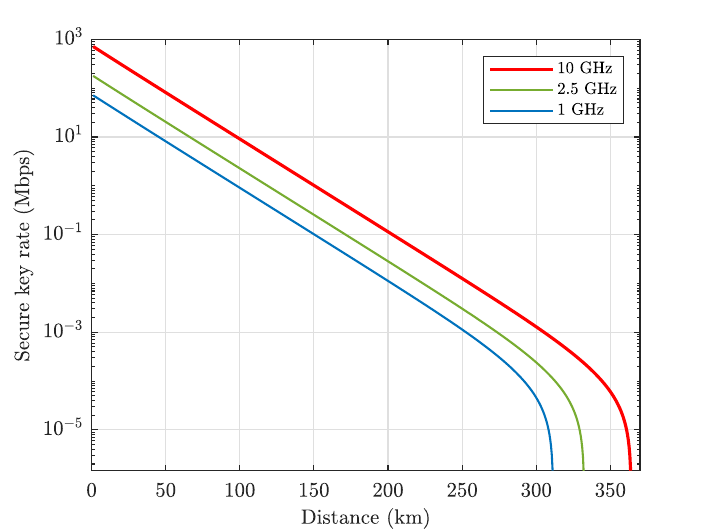}
\caption{Secure Key rate (Mbps) v.s. distance (km) for different system repetition frequencies. The proposed scheme achieves a higher SKR under the same distance condition and extends the transmission distance to more than 350 km.}
\label{fig12}
\end{figure}

Our simulations have demonstrated the superior performance of high-speed polarization modulation in long-distance QKD. We simulated the security key rate curve for the decoy state protocol based on the parameters in Tab.~\ref{tab1} \cite{fung2010practical,zhang2017improved,wang2005decoy}. $n_z$ is the finite block size. $u_s$ and $u_d$ are the signal and decoy intensity. $u_v$ is the vacuum intensity. $P_s$ and $P_d$ is the probability to choose the signal and decoy intensity, respectively. $P_Z$ is the probability to choose the $Z$ basis. $f_{EC}$ is the efficiency of error correction in the post-processing procedure. $e_d$ is the total polarization measurement error of system. $\alpha$ is the loss of fiber. $\varepsilon_{sec}$ is the security parameter. The simulation results for system repetition frequencies of 1 GHz, 2.5 GHz, and 10 GHz are shown in Fig.~\ref{fig12}. The proposed scheme achieves a higher SKR under the same distance condition. Specifically, the scheme extended the transmission distance of QKD beyond 350 kilometers.

\begin{table}[h]
\caption{Parameters used in the simulation}\label{tab1}%
\setlength{\tabcolsep}{3mm}
\begin{tabular}{@{}lllll@{}}
\midrule
$P_d=0.25$  & $P_s=0.5$ & $P_z=0.5$   & $n_z=10^8$\\
$u_d=0.11$ & $u_s=0.54$ & $u_v=0.35$   & $e_d=0.5\%$\\
 $f_{EC}=1.05$ & $\varepsilon_{sec}=10^{-10}$ & $\alpha=0.19$ dB/km\\
\botrule
\end{tabular}
\end{table}

\section{Conclusions}\label{sec5}

We eliminate the strong reverse modulation present in the self-compensating optics by utilizing the non-reciprocity of LiNbO$_3$ modulators in the RF band and achieve robust, low-error and high-speed polarization modulation. Our proposed encoder maintains high robustness while the system repetition frequency is no longer limited by the optical structure. It can be built by all commercially available components and does not introduce shortcomings such as high half-wave voltages. The measured average intrinsic quantum bit error rate of different polarization states is 0.53$\%$ over 10 min under the system repetition frequency of 10 GHz without any compensation. The simulation results prove that the proposed scheme achieves a higher SKR under the same distance condition and extends the transmission distance to more than 350 km. This work can maintain long-distance communication at high secure key rate, which is essential for the development of safe and reliable satellite-based communication technologies.

\subsubsection*{Abbreviations}

\begin{tabular}{@{}ll}
QKD & quantum key distribution \\
SKR & secure key rate \\
QBER & quantum bit error \\
PMD & polarization mode dispersion \\
PM & phase modulator \\
LiNbO$_3$ & lithium niobate \\
RF & radio-frequency \\
BS & beam splitter \\
PBS & polarization beam splitter \\
PMF & polarization-maintaining fiber \\
FWHM & full width at half maximum \\
IQBER & intrinsic quantum bit error \\
PER & polarization extinction ratio \\
\end{tabular}

\section*{Declarations}

\backmatter

\bmhead{Availability of data and materials}

The source data are available from the corresponding author upon reasonable request.

\bmhead{Authors’ contributions}

Y. Dai, B. Liu and H. Xu guided the project. Z. Wang and J. Li conceived the idea and designed the experiment. Z. Wang, H. Han and J. Huang performed the experiments and measurements. C. Wang, P. Zhang and F. Yin discussed the results. Z. Wang wrote the manuscript with the help of Y. Dai, B. Liu and H. Xu. K. Xu supervised the project. All authors read and approved the final manuscript.

\bmhead{Competing interests}

The authors declare no competing interests.

\bmhead{Funding}

This work was supported by the Science and Technology Innovation Program of Hunan Province (2023RC3003) and the National Natural Science Foundation of China (62071055, 62135014).

\bmhead{Ethics approval and consent to participate}

Not applicable.

\bmhead{Acknowledgements}

Not applicable.

\bmhead{Consent for publication}

All authors agree with the publication.



\begin{thebibliography}{10}
\providecommand{\doi}[1]{\url{https://doi.org/#1}}
\bibcommenthead

\bibitem[\protect\citeauthoryear{Xu et~al.}{2020}]{xu2020secure}
Xu F, Ma X, Zhang Q, Lo HK, Pan JW.
\newblock Secure quantum key distribution with realistic devices.
\newblock Reviews of modern physics. 2020;92(2):025002.

\bibitem[\protect\citeauthoryear{Liao et~al.}{2017}]{2017Satellite}
Liao SK, Cai WQ, Liu WY, Zhang L, Pan JW.
\newblock Satellite-to-ground quantum key distribution.
\newblock Physical Review Letters. 2017;.

\bibitem[\protect\citeauthoryear{Gr{\"u}nenfelder
  et~al.}{2018}]{grunenfelder2018simple}
Gr{\"u}nenfelder F, Boaron A, Rusca D, Martin A, Zbinden H.
\newblock Simple and high-speed polarization-based QKD.
\newblock Applied Physics Letters. 2018;112(5).

\bibitem[\protect\citeauthoryear{Li et~al.}{2023}]{li2023high}
Li W, Zhang L, Tan H, Lu Y, Liao SK, Huang J, et~al.
\newblock High-rate quantum key distribution exceeding 110 Mb s--1.
\newblock Nature photonics. 2023;17(5):416--421.

\bibitem[\protect\citeauthoryear{Gr{\"u}nenfelder
  et~al.}{2020}]{grunenfelder2020performance}
Gr{\"u}nenfelder F, Boaron A, Rusca D, Martin A, Zbinden H.
\newblock Performance and security of 5 GHz repetition rate polarization-based
  quantum key distribution.
\newblock Applied Physics Letters. 2020;117(14).

\bibitem[\protect\citeauthoryear{Xu and Wang}{2020}]{xu2020intrinsic}
Xu H, Wang S.
\newblock An intrinsic-stabilization polarization encoder for quantum key
  distribution.
\newblock In: Sixth Symposium on Novel Optoelectronic Detection Technology and
  Applications. vol. 11455. SPIE; 2020. p. 1359--1362.

\bibitem[\protect\citeauthoryear{Stein et~al.}{2023}]{stein2023robust}
Stein A, L{\'o}pez~Grande IH, Castelvero L, Pruneri V.
\newblock Robust polarization state generation for long-range quantum key
  distribution.
\newblock Optics Express. 2023;31(9):13700--13707.

\bibitem[\protect\citeauthoryear{Agnesi et~al.}{2019}]{agnesi2019all}
Agnesi C, Avesani M, Stanco A, Villoresi P, Vallone G.
\newblock All-fiber self-compensating polarization encoder for quantum key
  distribution.
\newblock Optics letters. 2019;44(10):2398--2401.

\bibitem[\protect\citeauthoryear{Luo et~al.}{2022}]{luo2022intrinsically}
Luo W, Li Y, Li Y, Tao X, Han L, Cai W, et~al.
\newblock Intrinsically stable 2-ghz polarization modulation for
  satellite-based quantum key distribution.
\newblock IEEE Photonics Journal. 2022;14(5):1--6.

\bibitem[\protect\citeauthoryear{Muller et~al.}{1997}]{muller1997plug}
Muller A, Herzog T, Huttner B, Tittel W, Zbinden H, Gisin N.
\newblock “Plug and play” systems for quantum cryptography.
\newblock Applied physics letters. 1997;70(7):793--795.

\bibitem[\protect\citeauthoryear{Boaron et~al.}{2018}]{boaron2018simple}
Boaron A, Korzh B, Houlmann R, Boso G, Rusca D, Gray S, et~al.
\newblock Simple 2.5 GHz time-bin quantum key distribution.
\newblock Applied Physics Letters. 2018;112(17).

\bibitem[\protect\citeauthoryear{Li and Herczfeld}{2006}]{li2006novel}
Li Y, Herczfeld PR.
\newblock Novel attenuation-counter-propagating phase modulator for highly
  linear fiber-optic links.
\newblock Journal of lightwave technology. 2006;24(10):3709--3718.

\bibitem[\protect\citeauthoryear{Fung et~al.}{2010}]{fung2010practical}
Fung CHF, Ma X, Chau H.
\newblock Practical issues in quantum-key-distribution postprocessing.
\newblock Physical Review A—Atomic, Molecular, and Optical Physics.
  2010;81(1):012318.

\bibitem[\protect\citeauthoryear{Zhang et~al.}{2017}]{zhang2017improved}
Zhang Z, Zhao Q, Razavi M, Ma X.
\newblock Improved key-rate bounds for practical decoy-state
  quantum-key-distribution systems.
\newblock Physical Review A. 2017;95(1):012333.

\bibitem[\protect\citeauthoryear{Wang}{2005}]{wang2005decoy}
Wang XB.
\newblock Decoy-state protocol for quantum cryptography with four different
  intensities of coherent light.
\newblock Physical Review A—Atomic, Molecular, and Optical Physics.
  2005;72(1):012322.

\end{thebibliography}
\end{document}